# Real-Time Device Reach Forecasting Using HLL and MinHash Data Sketches

Chandrashekar Muniyappa
Samsung Electronics America
Mountain View, CA, USA
e-mail: c.muniyappa@samsung.com

Kendall Willets
Samsung Electronics America
Mountain View, CA, USA
e-mail: k.willets@samsung.com

Sriraman Krishnamoorthy
Samsung Electronics America
Mountain View, CA, USA
e-mail: sriraman.k@samsung.com

*Abstract*—Predicting the right number of TVs (Device Reach) in real-time based on a user-specified targeting attributes is imperative for running multi-million dollar ADs business. The traditional approach of SQL queries to join billions of records across multiple targeting dimensions is extremely slow. As a workaround, many applications will have an offline process to crunch these numbers and present the results after many hours. In our case, the solution was an offline process taking 24 hours to onboard a customer resulting in a potential loss of business. To solve this problem, we have built a new real-time prediction system using MinHash and HyperLogLog (HLL) data sketches to compute the device reach at runtime when a user makes a request. However, existing MinHash implementations do not solve the complex problem of multilevel aggregation and intersection. This work will show how we have solved this problem, in addition, we have improved MinHash algorithm to run 4 times faster using Single Instruction Multiple Data (SIMD) vectorized operations for high speed and accuracy with constant space to process billions of records. Finally, by experiments, we prove that the results are as accurate as traditional offline prediction system with an acceptable error rate of 5%.

## ABBREVIATIONS AND ACRONYMS

HLL (HyperLogLog), MinHash, Hypercubes (multidimensional array of data), Campaign (Order placed by an advertiser in an Ad exchange, includes Placement and Creative Objects), Placement (an Object that holds targeting dimensions and Creative Objects specified by the advertiser), Creative (an object that holds targeting dimensions and the actual graphical image or video of an AD displayed on the screen), ETL (Extraction, Transformation, and Loading of data), SIMD (Single Instruction Multiple Data), EMR (Amazon Elastic MapReduce clusters), S3 (Amazon Simple Storage Service).

## I. INTRODUCTION

In the Ad business, advertisers will configure the targeting attributes like device type and year, programs and channels watched, user country, age and language preferences to name a few, at both placement and creative levels as include (only include the population that satisfies the selected targeting dimensions) and exclude (consider the population that compliments the selected targeting dimensions) criteria [14] and would like to know how many audiences they can reach. In Table I, one can see the targeting dimensions that are currently supported in the system, each dimension consists of at least 20 attributes. These are unique record counts only the actual raw dataset size is at least 5 times larger than these counts.

TABLE I. TARGETING DIMENSIONS

| Targeting Dimensions | # Unique records |
|---|---|
| DeviceProfile | 248,051,392 |
| Program | 47,733,704 |
| Channel | 42,408,271 |
| AppUsage | 137,093,718 |
| TasteGraph | 27,260,445 |
| CampaingExposure | 107,206,018 |
| DataSegment | 154,469,481 |
| CustomTargeting | 51,927,228 |
| ConnectedDevices | 40,680,174 |
| DemographicTargeting | 26,023,376 |

Traditional SQL joins on billions of records across multiple targeting dimensions will take a lot of time to compute the results. Therefore, in the current system predicting device reach is an offline process, where advertisers have to wait for 24 hours for the algorithms to complete and then make a decision, if they decide to make any changes then again, they have to wait for another 24 hours to see the effect of changes. This has adverse effects on business and revenue. To solve this crucial problem, we have built this new real-time forecasting system, which will provide real-time estimates to users as and when they modify the targeting attributes and help them in making the business decisions instantaneously.

Forecasting plays a pivotal role in the Ad booking system as advertisers and publishers will make the budget decisions and sign the contract based on the forecasting numbers. Hence, any kind of over or under predictions will result in a revenue loss for publishers. Therefore, speed and accuracy are important in forecasting.

Section II will talk about the background, Section III will talk about the approach we have taken, Section IV will present the experiments and results, and finally, Section V will cover conclusion and future work.

## II. BACKGROUND

Prediction systems can be developed in multiple ways using AI and statistical modelling techniques; however, they are complex, difficult to implement and maintain, and error-prone [1, 2, 3]. On the contrary, there is a family of stochastic algorithms that can be used for forecasting [4], they are gaining popularity [15] as they are fast, efficient, and have permissible error rates. In this work, we will show how HLL and MinHash algorithms can be used to forecast the device reach. This technique works on any dataset and the solution can be easily extended to build a forecasting system in any domain.

HyperLogLog [5] – HLL is mainly used to estimate the cardinality of a stream with constant space. We can adjust the size of memory to keep the error-rate less than 2%. Furthermore, the algorithm can be modified for better performance and accuracy [6]. However, by design HLL can be used to determine only the cardinality of a population, to obtain the intersection one can use inclusion-exclusion principle [7] but it is a cumbersome process and has a high error rate [4, 8, 9]. To overcome these problems, MinHash is used to find the intersection.

MinHash [10] is based on Locality Sensitive Hashing mainly used to determine how similar the given sets are based on the Jaccard similarity. Many latest DBs [15] provide these algorithms as a built-in feature, but the problem is they do not expose signatures that can be broadcasted over a network, so that, the signatures can be held in memory of an online application. There are open-source libraries [16] to solve this problem. However, none of the existing solutions support multilevel or nested set aggregation feature which was crucial for our use case. This work will show how we have solved this problem. The contributions of this work are as follows:

- Improved the MinHash algorithm to run 4 times faster by applying SIMD vectorized operations [17].
- We have enhanced the open-source library [16] to support multilevel set aggregation, so that, the MinHash algorithm can work on both original MinHash signatures and also on the intermediate MinHash signatures representing the Jaccard similarity between sets.

In our discussion, set and subsets represent Hypercube records [11]. Union and Intersection refers to the HLL cardinality and MinHash intersection (Jaccard Similarity) functions. The HLL and MinHash data sketches are custom MapReduce UDAFs (User Defined Aggregate Functions) installed on Vertica columnar DB [12].

## III. APPROACH

The work flow has the following two main steps:
- Generating the sketches
- Querying the sketches

### A. Generating the Sketches

Apply Extraction, Transformation, and Loading (ETL) techniques to transform the data as per the business requirements. This is the first step in the estimation process, as part of this step, targeting dimensions will be fetched from S3 at scheduled intervals and will be transformed based on the business rules using spark jobs [13]. At the end of this step, the data stored on S3 will be loaded into Vertica DB temporary tables, an example of a simplified version of the DeviceProfile dataset is shown in the Table II.

TABLE II. DEVICEPROFILE RAW DATASET

| PSID | country | year | … | chipset |
|---|---|---|---|---|
| 12abc3 | US | 2011 | … | KRM |
| ef1268 | US | 2011 | … | KRM |
| 980jkjk | CA | 2018 | … | CHM |
| kasbudu | CA | 2018 | … | CHM |
| kasbudu | CA | 2019 | … | CHM |

Next step is to classify or aggregate the data into different subsets, using HLL and MinHash data sketches. We will partition a set (Targeting dimension) containing billions of records into many subsets, based on the business rules of a given set. The Pseudo Identifier - PSID unique device ID (64-bit hash, MAC address) of a TV for each record in a subset will be hashed using HLL and MinHash sketches to reduce entire subset into a single aggregated record. We will compute include and exclude sets signatures for each subset, where an include signature will give the population of the devices which belong to the subset and exclude signature will give the complement of a given include-set. The process will be repeated for each of the subsets and the final aggregated set (Hypercube) will have one record (base cuboid) for each subset with include and exclude HLL and MinHash hash values. The hypercubes generated like this will be stored in the Vertica analytics DB. Please find the example Table III, which demonstrates this process, for the input Table II from the previous step.

TABLE III. HLL AND MINHASH SINGNATURES

| country | year | chipset | hll | exhll | minhash | exminash |
|---|---|---|---|---|---|---|
| US | 2011 | KRM | xyx | wew | oiu | poip |
| CA | 2018 | CHM | yyy | fgh | sdf | kljs |
| CA | 2019 | CHM | ppp | jkj | asd | mnb |

As one can see from the example Table III, the aggregation process using group by clause on the country, year and chipset attributes will result in 3 subsets with PSIDs converted to include and exclude HLL and MinHash binary signatures (hash values). In the actual production environment, we will be grouping on around 20 attributes and millions of records will be aggregated and reduced to thousands of summarized records (base cuboids) like above.

To compute the include-set signatures (HLL and MinHash columns) for a given targeting dimension, we will aggregate a targeting dimension into subsets based on the given set of attributes using SQL group by operator and aggregate the PSIDs in each subset by calling the HLL and MinHash libraries. However, to construct the exclude-set signatures (exhll and exminhash columns) we need to find the complement for each of the given include-set with respect to the device universe (All the TVs for a given country). This is computationally intensive because we will not have join columns to join the dimensions. Therefore, the traditional cross join approach will result in Trillions of records to be processed, for example, the device universe that represents the number of active TVs in the US alone will be 40 Million and the number of TVs that watched a program will be around 200K, the cross product of this will be around 8 Trillion records. The cross-join query will take around 20 hours to complete, on a 6 node, 64GB EMR cluster. To solve this problem a novel taxonomy-based query to generate the exclude-set signatures was invented that reduced the run time to an hour [18].

The original MinHash library [16] is enhanced with Intel intrinsics AVX2 and AVX512 SIMD instructions [19], where the traditional looping, loading, and storing code is replaced with the SIMD instructions [20]. This reduced the runtime by 4 times, we ran multiple rounds of experiments over several data points and found that the runtime gain was consistent across the runs, as shown in Table IV.

TABLE IV. 4X MINHASH RUNTIME IMPROVEMENT

| Before | After |
|---|---|
| 2.45 Seconds | 0.599 Seconds |

The ETL pipelines are scheduled to run hourly, daily, and on-demand basis to build these hypercubes consisting of latest HLL and MinHash signatures.

### B. Querying the Sketches

The advertiser will set up a campaign with the preferred targeting dimensions at placement and creative levels and submit a request to get an estimate, once he is satisfied with the setup, the placement will be approved to deliver. The Fig(2) in the appendix section shows an approved live placement delivering the Ads, Fig(3) shows placement level targetings and a creative assigned to it, Fig(4) shows the details of a creative, and finally, Fig(5) shows all the targeting dimensions that can be added at placement and creative levels.

Upon receiving the request, the server will translate it into a dynamic SQL query to query the include and exclude HLL and MinHash sketches of the given targeting dimension hypercubes. A placement can have multiple targetings and creatives assigned with both include and exclude criteria. To compute the estimation, first similarity is measured between all the placement level targetings and then the similarity between the creative level targetings are measured, if there is more than one creative then the similarity measurements of all the creatives are unioned to form one single aggregated creative level similarity measurement. Finally, the similarity between the two intermediate signatures is measured. For example, if P1 is the Placement, with Creatives C1, C2, C3 ... CN and has the targeting attributes T1, T2, T3, ... TN, and say each of the Creatives have targeting attributes CT1, CT2, CT3, ...CTN assigned then we compute the estimation as follows.

Select hllest(hllagg(hll or exhll)) as cardinality * mhjaccard(mhagg(minhash or exminhash)) as jaccardRatio
FROM
((P1(T1 ∩T2 ∩ … ∩ TN )) ∩
((C1(CT1 ∩ CT2 ∩ … ∩ CTN )) ∪
((C2(CT1 ∩ CT2 ∩ … ∩ CTN )) ∪ … ∪
((CN(CT1 ∩ CT2 ∩ … ∩ CTN )))))

where hllest, hllagg, mhagg, and mhjaccard are UDAFs developed using C programming language and installed on a Vertica DB. The above computation can be pictorially represented as shown in the Fig (1).

Existing MinHash algorithm implementations does not support multilevel aggregation or intersection. They will only accept first level signatures and not the intermediate signature of the common elements across sets as shown in Fig (1). To solve this problem, first, we need to capture the signature of common elements (Jaccard similarity signature) across sets this is done as shown in the code listing 1 in the appendix section. The function is registered as mhjaccard UDAF in Vertica DB, once the function completes we can save the Jaccard signatures in a DB as shown in Table III. The final Jaccard signatures are used to compute the Jaccard ratio, by counting the number of bits set in the Jaccard bitmask and dividing it by the number of bins (length of the bitmask).

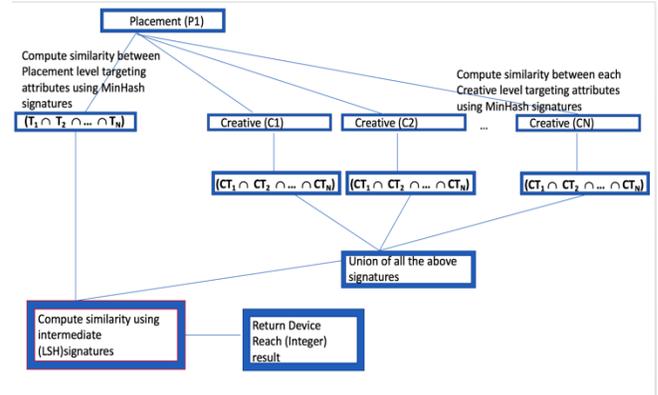

Figure 1. Similarity Based on Intermediate Signatures.

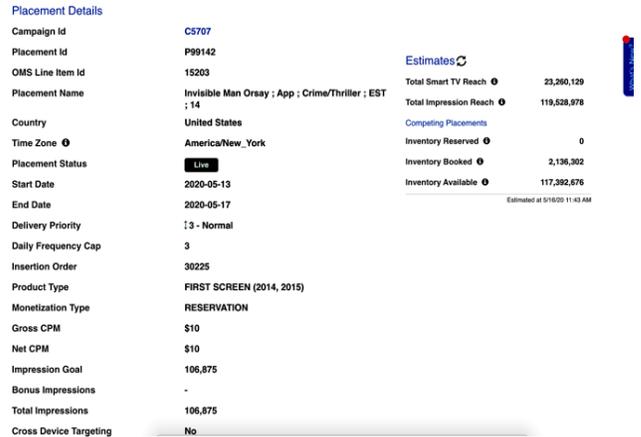

Figure 2. Placement.

Figure 3. Placement Level Targetings.

Figure 4. Creative.

Figure 5. Targeting dimensions.

Once we have the Jaccard ratio, one can compute the intersection by multiplying the Jaccard ratio with the cardinality of a set provided by HLL

$$DeviceReach = JaccardRatio * HLLCardinality \quad (1)$$

This is derived from

$$JaccardRatio = |A \cap B| / (|A| + |B| - |A \cup B|)$$

$$|A \cap B| = JaccardRatio * (|A| + |B| - |A \cup B|) \quad (2)$$

IV. EXPERIMENTS AND RESULTS

The Table V shows the device reach forecasted and the time taken for a single Placement with different combinations of include and exclude criteria at Placement and Creative level targetings.

TABLE V. FORECATSED VALUES AND TIME TAKEN

| #Placement Level Taregtings | #Creatives | #Creative Level Targetings | Forecasting | Forecasting Time in Seconds |
|---|---|---|---|---|
| 5 | 0 | 0 | 5,809,483 | 4.6 |
| 5 | 1 | 5 | 5,625,236 | 5.4 |
| 10 | 1 | 10 | 6,389,770 | 5.2 |
| 10 | 5 | 30 | 17,221,260 | 5.6 |

With this new approach, the system is forecasting device reach in few seconds when compared to the existing system which takes 24 hours. As the predictions are real-time, business is now able to make the decisions instantaneously and sign the contracts without having to wait for 24 hours, resulting in, doubling the revenue and expanding the business to new countries.

To measure the accuracy of the forecasted values, we can compute the relative error rate as follows:

$$(|Truevalue - Observedvalue| / Truevalue) * 100.$$

TABLE VI. ACCURACY

| True value from SQL | Predicted value | Error rate |
|---|---|---|
| 5,803,033 | 5,809,483 | 0.111 |
| 6,650,830 | 6,389,770 | 3.925 |
| 16,850,470 | 17,221,260 | 2.2 |

We ran the experiments on several data points in the production environment and computed the accuracy the error rates were less than 5%. A sample of this is shown in Table VI.

V. CONCLUSION AND FUTURE WORK

As shown in this work, we have enhanced the MinHash algorithm implementation to run 4 times faster and to handle multilevel aggregations, and have successfully applied to forecast the device reach in real-time with error-rate less than 5%. Now, the business can make immediate decisions and onboard new advertisers for higher revenue growth. However, this approach works only with non-seasonal data, as part of the future work one can enhance the algorithm to work with seasonal data by adding seasonality factor to the objective function.

APPENDIX: MULTILEVEL SET AGGREGATION C CODE

```c
void mh_jaccard (inputMinash){
if(bins is empty){
   init(bins, inputMinash)
} else{
__m128i const a=_mm_load_si128 (inputMinhash); __m128i const b=__m_load_si128(bins);
__m128i const eq = _mm_cmeq_epi32(a,b);
__m128i const common =__mm_store_si128(bins,eq);

uint32_t *ucommon = (uint32_t *)&common;
uint32_t int_values[4]={0,0,0,0};
```

```
uint32_t *pa =((uint32_t *) inputMinash.data())+4*i;
uint32_t *pb = ((uint32_t *) bins.data()) + 4*i;

 for(int j = 0; j < 4; j++){
     if( ucommon[j] == -1){
if(pb[j] == 0) {
         ubins[j] = 0;
     }
     int_values[j]=pb[j];
   }}

_mm_store_si128(_m + i, cum);
_mm_store_si128(_mIntersect + i,
_mm_set_epi32(int_values[3], int_values[2],int_values[1] , int_values[0])); }
```